# Noncollinear Antiferromagnetic Spintronics

*Hongyu Chen, Peixin Qin, Han Yan, Zexin Feng, Xiaorong Zhou, Xiaoning Wang,*

*Ziang Meng, Li Liu, Zhiqi Liu\**

School of Materials Science and Engineering, Beihang University, Beijing 100191, China.

## Abstract

Antiferromagnetic spintronics is one of the leading candidates for next-generation electronics. Among abundant antiferromagnets, noncollinear antiferromagnets are promising for achieving practical applications due to coexisting ferromagnetic and antiferromagnetic merits. In this perspective, we briefly review the recent progress in the emerging noncollinear antiferromagnetic spintronics from fundamental physics to device applications. Current challenges and future research directions for this field are also discussed.

Corresponding author: zhiqi@buaa.edu.cn

Spintronics,[1–3] which focuses on the physics and exploitation of the intrinsic spin of electrons in addition to the charge, has revolutionized information technologies and holds the potential for "more-than-Moore" electronics.[4] For the past decades, spintronics has been dominated by ferromagnets where the magnetic order permits spontaneous magnetization. In contrast, antiferromagnets, in which no net moment exists due to fully compensated magnetic structures, have been playing an auxiliary role in spintronic devices until recently. In 2016, the breakthrough demonstration of a room-temperature antiferromagnetic CuMnAs-based memory[5] reveals that antiferromagnets can act as the core of spintronic devices as well. Moreover, compared to ferromagnets, antiferromagnets are endowed with vanishing stray fields and faster spin dynamics, which can enable higher packing densities and superior response frequencies of up to terahertz (THz) for information devices, respectively.[6–8] Therefore, antiferromagnetic spintronics has been regarded as one of the leading candidates for next-generation electronics and has attracted surges of interest.[9,10]

Hitherto, two categories of antiferromagnets have been found promising for realizing antiferromagnet-centered spintronic devices. The first kind includes certain collinear antiferromagnets with locally broken inversion symmetry that permits Néel spin-orbit torques to rotate the Néel vectors,[11] such as tetragonal CuMnAs[5,12] and Mn$_2$Au.[13,14] The second class comprises some coplanar noncollinear antiferromagnets, *i.e.*, antiferromagnets with coplanar noncollinear magnetic structures, such as hexagonal $D0_{19}$-type Mn$_3$$X$ ($X$ = Sn, Ge, Ga),[15–17] cubic Mn$_3$$X$ ($X$ = Ir, Pt, Rh),[18–20] and antiperovskite Mn$_3$$A$N ($A$ = Ga, Sn, Ni).[21–23] Owing to the special noncollinear antiferromagnetic order and the resultant unique symmetry, these materials not only inherit most inherent merits of antiferromagnets, but also resemble ferromagnets in many aspects, which makes them both intriguing in physics and viable for application. In this perspective, we briefly review the recent progress in the emerging noncollinear antiferromagnet-based spintronics, which we term as "noncollinear antiferromagnetic spintronics" and illustrate in Figure 1, from fundamental physics to device applications. In addition, existing challenges are outlined and possible future research directions are envisioned.

# Noncollinear Antiferromagnetic Spintronics

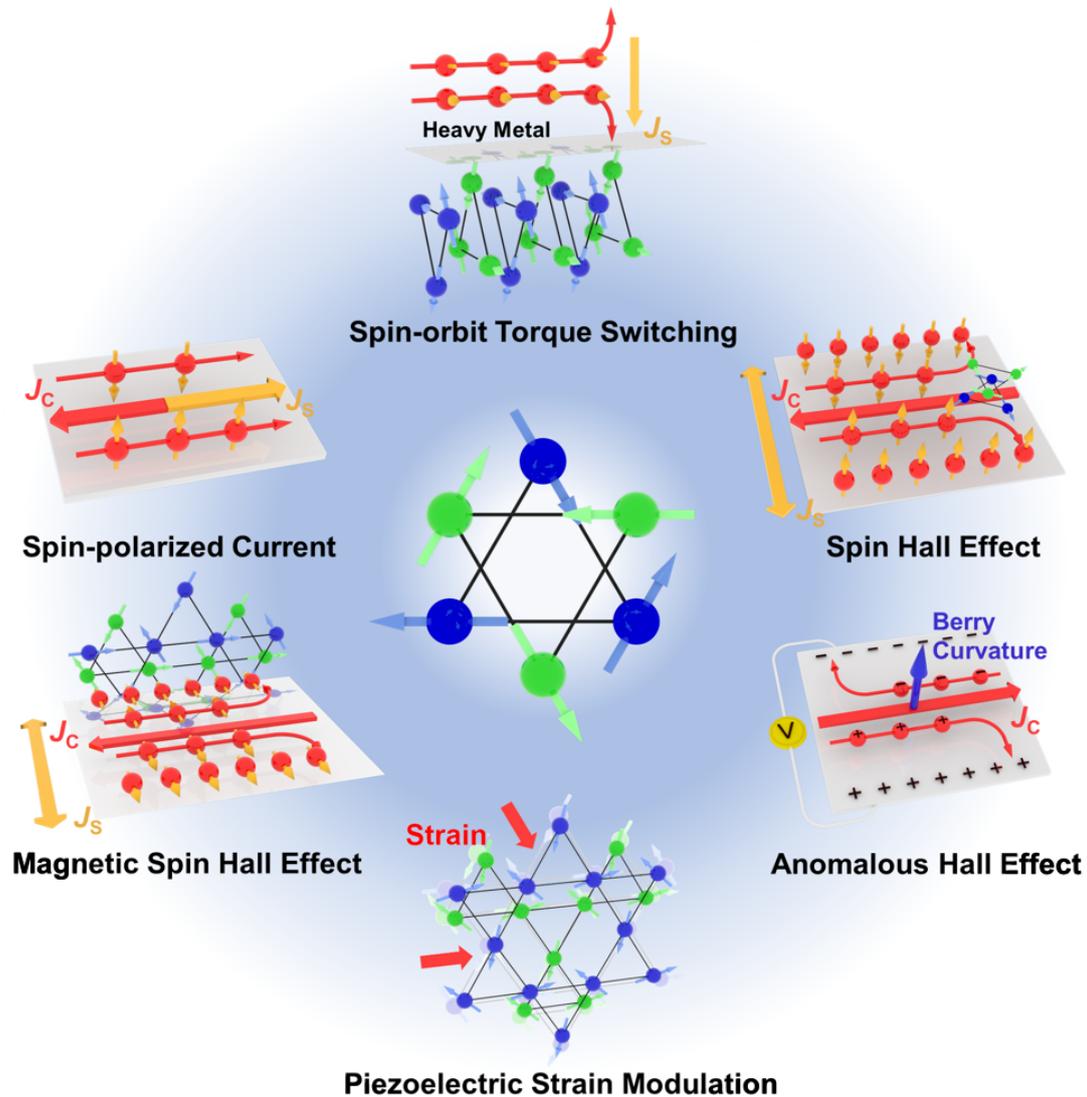

**Fig. 1** Conceptual schematics for some major exotic physical phenomena and spin manipulation methods relevant to noncollinear antiferromagnetic spintronics.

Basically, the presence of the long-range triangular antiferromagnetic order (the center of Figure 1) breaks the macroscopic time-reversal symmetry, which is typically broken in ferromagnets, of noncollinear antiferromagnets. Accordingly, this, together with spin-orbit coupling (SOC), gives rise to non-vanishing $k$-space Berry curvature and a consequent large anomalous Hall effect (AHE), as is the case with ferromagnets.[24] The AHE in noncollinear antiferromagnets was theoretically predicted in 2014[25,26] and soon experimentally verified in $Mn_3Sn$ in 2015 (Figure 2).[27] Despite the negligibly small magnetic moment of a few m$\mu_B$ per Mn due to spin canting,[28] the anomalous

Hall conductivity of bulk single-crystal $Mn_3Sn$ reaches ~20 $\Omega^{-1}$ $cm^{-1}$ at room temperature, comparable to those of certain ferromagnetic metals.[29] Such a discovery conflicts the conventional wisdom that the AHE is proportional to magnetization, and further corroborates the topological nature of the intrinsic AHE–Berry curvature.[24] Up to now, the AHE has been experimentally found in most of the aforementioned noncollinear antiferromagnets.[30–41] In addition, the thermal counterpart of the AHE, *i.e.*, the anomalous Nernst effect, has been observed in the meantime.[42–45]

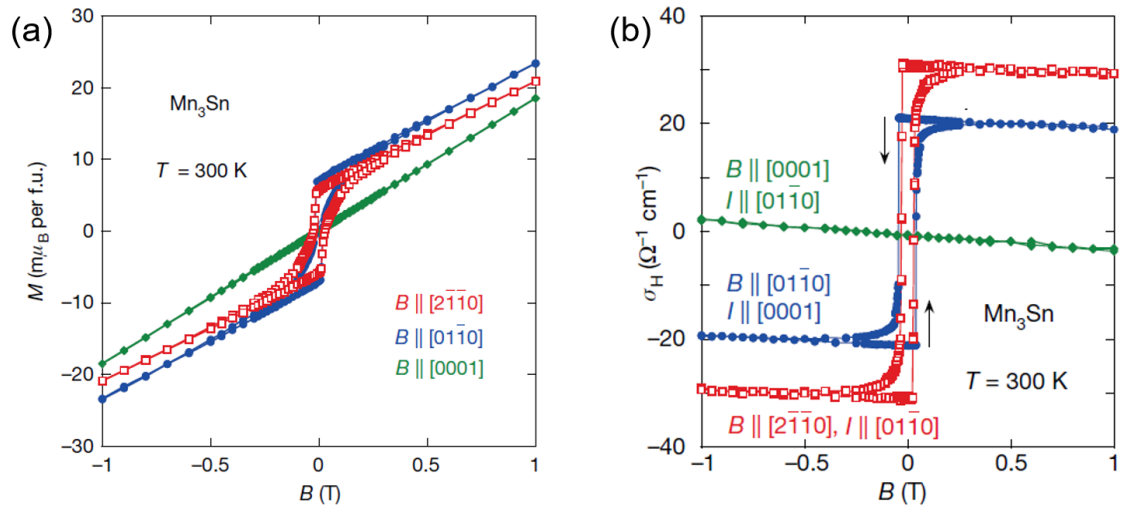

**Fig. 2** (a) The magnetization (*M*) curves and (b) the Hall conductivity ($\sigma_H$) versus the applied field (*B*) at 300 K of $Mn_3Sn$ measured in *B* || [2$\bar{1}\bar{1}$0], [01$\bar{1}$0] and [0001].[27] Copyright 2015, Springer Nature.

Based on the same symmetry consideration, spin-polarized currents, another inherent attribute of ferromagnets, are expected to exist in noncollinear antiferromagnets as well. The prediction of the spin-polarized currents in noncollinear antiferromagnets was made by Železný *et al.* in 2017.[46] It was theoretically found that the spin polarization of the electrical currents in $Mn_3X$ compounds could be comparable to that of ferromagnets even in the absence of SOC. Furthermore, distinct from ferromagnets, the spin texture near the Fermi level induced by the triangular magnetic structures can result in both longitudinal spin-polarized currents and transverse net spin currents upon applying electric fields. It should be emphasized that such an effect is fundamentally different from the ordinary spin Hall effect (SHE) originating from SOC,[47] for the former is odd under time reversal (*T*), while the latter is even under *T*. Recently, similar

results have been obtained for Mn$_3$*A*N.[48]

Although little experimental evidence for the predicted longitudinal spin-polarized currents has been put forward up to now, they indeed suggest a feasible antiferromagnetic version of the building blocks of conventional spintronics, such as spin-transfer torques (STTs), the giant magnetoresistance resistance (GMR) effect, and the tunneling magnetoresistance (TMR) effect. Very recently, these phenomena have been theoretically studied in detail,[49,50] which we will introduce later.

On the other hand, the *T*-odd transverse spin currents could be responsible for the recently discovered magnetic spin Hall effect (MSHE).[51] For the MSHE of Mn$_3$Sn, it was found that the polarization of the accumulated spins at bulk surfaces due to applied electrical currents changes its sign after reversing the triangularly ordered moments via magnetic fields (Figure 3). It is interesting to notice that such an abnormal effect phenomenally resembles the *T*-odd transverse spin currents proposed by Železný *et al*.[46] Regarding the intrinsic difference between the *T*-even ordinary SHE originating from SOC and the novel *T*-odd magnetic SHE, it was proposed that the *T*-odd MSHE is yielded by the linear response of inter-band spin density to an external electric field while the *T*-even ordinary SHE comes from the linear response of intra-band spin density to an external electric field.[51] As a result, the MSHE was predicted to exist in all single-crystal magnetic material systems with time-reversal-symmetry breaking. In addition, a recent study has ascribed the MSHE to the *T*-odd spin currents and provides a more intuitive interpretation with regard to the spin current vorticity in the Fermi sea.[52]

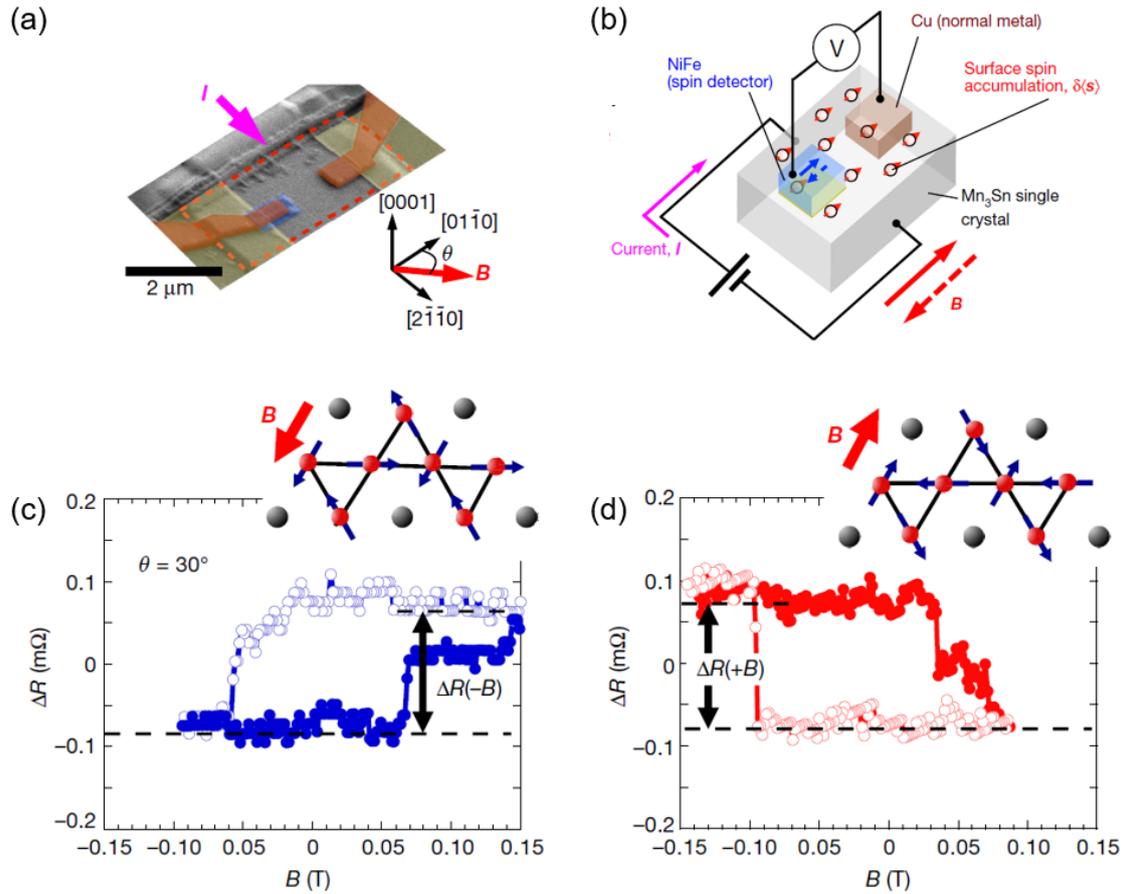

**Fig. 3** (a) Scanning electron microscope image of the spin-accumulation device. The red dashed line denotes the Mn$_3$Sn single crystal while the blue square is the ferromagnetic NiFe electrode. The non-magnetic Cu leads are indicated by the brown areas. (b) Schematic of the measurement geometry. The electrical current ($I$) was applied along [2$\bar{1}\bar{1}$0] while the external magnetic field ($B$) was applied within the basal plane of the device and the field angle $\theta$ was measured from [01$\bar{1}$0]. When the spin accumulation generated by the injected $I$ at Mn$_3$Sn surfaces has a component parallel to the NiFe magnetization, the electrochemical potential across the Mn$_3$Sn/NiFe interface is supposed to be changed and a voltage between the NiFe and Cu electrodes will be induced. (c), (d) Resistance ($\Delta R$) measured between the NiFe and Cu electrodes versus $B$ at room temperature. The Mn$_3$Sn had been magnetically saturated by a large $B$ of −0.75 T and +0.75 T before the measurement in (c) and (d), respectively. The insets show the corresponding magnetic order of Mn$_3$Sn; blue arrows represent Mn sublattice moments.[51] Copyright 2019, Springer Nature.

Despite the inexplicit underlying physics, the MSHE indicates a simple way to controlling the polarization of spin currents and thus the spin-orbit torques (SOTs) generated by noncollinear antiferromagnets–rotating external magnetic fields. This could be favorable for antiferromagnetic spintronics. Moreover, the accumulated spins in the MSHE have been shown to exhibit out-of-plane polarization, which can enable the energy-efficient field-free magnetization switching of perpendicularly magnetized

ferromagnetic layers.[53,54]

Apart from the ferromagnet-like aspects, the merits of antiferromagnets, such as the negligibly weak magnetic moments,[27,31] and the ability to serve as the pinning layer in magnetic junctions,[55–57] are gifted to noncollinear antiferromagnets as well. Particularly, they also possess the ultrafast spin dynamics due to antiferromagnetic exchange interactions that is favorable for high-speed data-possessing applications. This has been corroborated by the time-resolved cluster magnetic octupole oscillation of $Mn_3Sn$.[58] Notably, it was deduced that the effective damping of octupole dynamics can support an octupole switching timescale of less than 10 ps.

Additionally, the special magnetic order also endows noncollinear antiferromagnets with certain unique characteristics that are uncommon in general ferromagnets or antiferromagnets, such as the existence of the $T$-even SHE without SOC.[59] It has been shown that the special symmetry of noncollinear antiferromagnets can lead to large anisotropic spin Hall conductivity where the role of SOC could be completely replaced by the triangular magnetic structures, *i.e.*, the $T$-even SHE could survive in the absence of SOC.[60] Moreover, a recent study has revealed that SOC could even reduce the SHE.[61] The measured spin Hall angle of some $Mn_3X$ alloys can be comparable or even larger than Pt.[45,62–65]

More importantly, detailed analyses have revealed that the spin polarization in the SHE (as well as the Rashba-Edelstein effect) is not always orthogonal to the generated spin currents in noncollinear antiferromagnets due to their lowered symmetry by the triangular magnetic order.[62,66,67] Specifically, laterally broken magnetic mirror symmetry could give rise to out-of-plane spin polarization.[68–70] Recently, a more intuitive interpretation has been proposed based on the cluster magnetic octupole theory.[70] Notably, the geometry of such a phenomenon is exactly the same as the MSHE and thus it can be hard to distinguish them in practice. Perhaps a comparison of the (M)SHE before and after rotating the magnetic order could help. Nevertheless, regardless of the explicit physical origin, the out-of-plane spin polarization can lead to abnormal SOTs, as is the case with the MSHE. The field-free switching of perpendicularly magnetized ferromagnetic layers by the SOTs generated by noncollinear antiferromagnets has been experimentally demonstrated.[67,70]

As has been discussed above, most of the studies on noncollinear antiferromagnetic spintronics have been dedicated to fundamental physics up to now, indicating that this field is still at its infancy. In addition, it should be mentioned that some interesting phenomena, such as various magneto-optical effects,[36,71–75] which seem not straightforwardly relevant to spintronics are not covered in this Perspective. We guide interested readers to the references listed above. In the following paragraphs, we would like to introduce some recent progress that is beneficial for practical applications.

The basis to build spintronic devices lies in the effective control of magnetic spins. For ferromagnets, this can be readily achieved by magnetic fields or spin torques. However, such tactics are not amenable to simple collinear antiferromagnets due to fully compensated magnetic moments (except for the spin-flip and spin-flop transition under extremely large fields). On the other hand, thanks to the special magnetic order and the spin canting[28,76], one can manipulate the spins of noncollinear antiferromagnets in the same way as ferromagnets.[77] Indeed, the sign reversal of the AHE after reversing magnetic fields straightforwardly reveals the change of the magnetic order of noncollinear antiferromagnets. Nevertheless, generating magnetic fields via current coils is rather energy consuming and cannot support modern device applications.

In 2020, Tsai *et al*.[78] discovered that the triangular magnetic order of $Mn_3Sn$ can be manipulated by SOTs with a small in-plane field of ~0.1 T at room temperature (Figure 4). Specifically, for $Mn_3Sn$/heavy metal stacks, the applied longitudinal electrical currents can generate spin currents in the heavy metal, which are subsequently injected into the $Mn_3Sn$ layer and exert spin torques on the sublattice moments. When the applied longitudinal current density exceeds a critical value of $10^{10}$–$10^{11}$ A m$^{-2}$, the Hall voltage of $Mn_3Sn$ reverses its sign due to the rotation of the Weyl nodes resulted from the switched cluster-octupole polarization. In addition, the sign change of the AHE is determined by both the polarity of longitudinal currents and in-plane fields and the spin Hall angle of the heavy metal, completely analogous to the SOT switching scenario for ferromagnets.[79] Furthermore, the critical current density can be reduced to ~$10^7$ A m$^{-2}$ for an epitaxial sample,[80] a value comparable to those of ferromagnetic SOT devices.[79] Very recently, the SOT-induced octupole switching has been verified by the nitrogen-vacancy center technique that is able to qualitatively measure the local stray fields of Mn$_3$Sn at a sub-micrometer length scale.[81] In addition, similar switching signatures have been reported for Mn$_3$Ir and Mn$_3$GaN.[82–84] Although these discoveries are indeed exciting breakthroughs for noncollinear antiferromagnetic spintronics, the energy cost in the switching process is relatively high due to unavoidable Joule heating, which is an intrinsic shortage of spin torques.

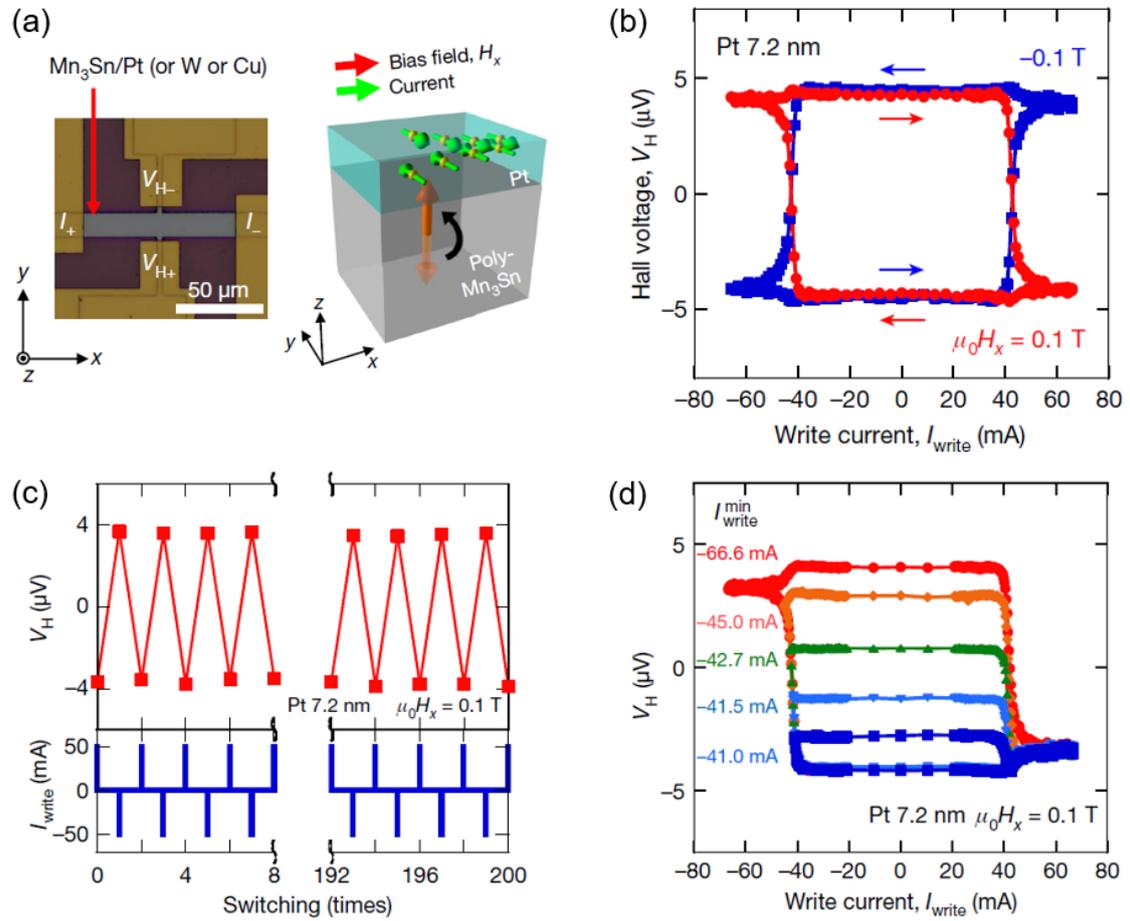

**Fig. 4** (a) The optical photo of the Mn$_3$Sn/nonmagnetic layer (Pt, W, or Cu) devices (left) and the schematic of spin-orbit-torque (SOT) switching (right). Write and read currents and the magnetic bias field ($\mu_0 H_x$) are applied along the *x* direction. The spin-polarized currents along the *z* direction (green arrows on yellow spheres) in Pt generated by the write current can exert SOTs on the cluster magnetic octupole (orange arrow) of Mn$_3$Sn and result in the switching of the polarization axis. (b) Hall voltage ($V_H$) versus write currents ($I_{write}$) for a Mn$_3$Sn/Pt (7.2 nm) device under $\mu_0 H_x = \pm 0.1$ T. (c) $V_H$ (top) and $I_{write}$ (bottom) for the same device measured at room temperature and under $\mu_0 H_x =$

0.1 T. $V_H$ measured by a read current of 0.2 mA changes its sign depending on the polarity of the $I_{write}$ pulse with a duration of ~100 ms. The switching of $V_H$ is performed for 200 times. (d) $I_{write}$-dependent $V_H$ for the same device measured at room temperature under $\mu_0 H_x = 0.1$ T. The magnitude of the switching of $V_H$ is affected by the minimum write current ($I_{write}^{min}$).[78] Copyright 2020, Springer Nature.

A more energy-efficient method to manipulate noncollinear antiferromagnetic spins is to utilize electric-field-generated strain.[6–8,85–88] Via constructing multiferroic heterostructures composed of noncollinear antiferromagnet/ferroelectric bilayer, the triangular magnetic order can be affected by the piezoelectric strain of the ferroelectric layer upon applying moderate electric fields, which can manifest as a variation in the AHE or magnetization loops. The core of such a tactic lies in the effective tuning of the competition between magnetoelastic energy and other magnetic anisotropy energy. Electric-field manipulation of noncollinear antiferromagnets has been demonstrated in $Mn_3X$[55–57,89,90] and $Mn_3AN$.[91–93]

Nevertheless, even though one takes no account of the energy cost, the switching timescales of the above two methods are limited to ~100 ps as they both rely on electric circuits.[94] In order to take full advantage of the ultrafast antiferromagnetic spin dynamics, *i.e.*, achieving a switching timescale on the order of ps, optical methods such as femtosecond laser pulses or the THz electric fields generated by femtosecond laser pulses could be employed, as is the case with ferrimagnets and collinear antiferromagnets.[95–106] Recently, a phased achievement has been made that the magnetic order of $Mn_3Sn$ thin films was found to be affected by high-power laser via the laser-generated heat in conjunction with weak magnetic fields, and the change in magnetic structure was evidenced by scanning thermal gradient microscopy, a method that utilizes the anomalous Nernst effect to image the magnetic domain.[107] However, the ultra-fast all-optical domain switching of noncollinear antiferromagnets remains unexplored.

Apart from the effective control of magnetic order that permits write-in, significant changes in physical responses induced by the variation of spins that allows reliable readout is another vital prerequisite for spintronic devices. There are two basic strategies to achieve large readout signals for noncollinear antiferromagnetic spintronic devices: utilizing the large AHE, or employing the antiferromagnetic magnetoresistance (MR) effects, including the anisotropic magnetoresistance (AMR), GMR, and TMR.

Currently, the sign-tunable AHE readout has been realized in the SOT devices (Figure 4c).[78,84,108] Also, multistate memristor-like behavior that is favorable for neuromorphic spintronics has been discovered (Figure 4d).[78] However, more efforts are needed to further enlarge the absolute change of the Hall voltage since the largest variation value obtained with a moderate current density is only several mV up to now.[108] Possible routes include improving sample quality and enhancing SOT efficiency. In addition, the AHE can also be effectively tuned or optimized by strain (pressure)[32,55–57,89,91–93,109,110] or chemical potential.[111]

One the other hand, a typical AMR ratio for an antiferromagnetic metal is only ~0.1% at room temperature,[5,112] which and is far from application and also holds for noncollinear antiferromagnets.[57] Although the MR ratio can be enhanced to ~10% via the anisotropic TMR effect of a $Mn_3Ga$/MgO/Pt tunneling junction,[56] it remains much smaller than the ferromagnetic TMR ratio of more than 200%.[113] Very recently, noncollinear antiferromagnetic GMR junctions composed of a nonmagnetic conducting layer sandwiched by two identical noncollinear antiferromagnetic layers have been proposed.[49] It was shown that the STTs in such junctions can enable a deterministic switching on a picosecond timescale. Moreover, the GMR ratio could be even larger than that for ferromagnets and is insensitive to disorder, which is favorable for laboratory demonstration.

Aside from GMR junctions, the noncollinear antiferromagnetic TMR effect,[46,48] a concept that was put forward years ago, might be another promising route. Very recently, noncollinear antiferromagnetic TMR junctions, *i.e.*, an insulating tunneling barrier sandwiched by two identical noncollinear antiferromagnetic electrodes, have been theoretically studied in detail.[50] It was found that the spin polarization of the Fermi surfaces of noncollinear antiferromagnets are related to the direction of the Néel vector, which enables a TMR effect similar to the conventional ferromagnet-based junctions.

Furthermore, depending on the relative orientation of the Néel vectors of the antiferromagnetic layers, four nonvolatile resistance states can be obtained and the highest TMR ratio reaches ~300% for a $Mn_3Sn$-based junction with a vacuum barrier. Such exciting results are imperatively awaiting their experimental demonstration.

Aside from conventional spintronic devices, noncollinear antiferromagnets are also candidate materials for THz technologies. For example, it has been shown that $Mn_3Sn$ and $Mn_3Ir$ can be unutilized to construct THz emitters.[114,115] In addition, the large AHE signals have been shown to survive at a THz frequency at room temperature, which is beneficial for THz information reading (Figure 5).[116] Moreover, exploiting the ultrafast spin dynamics, noncollinear antiferromagnet-based THz oscillators have been proposed, which have potential applications in THz sensing, imaging, and neuromorphic computing.[117]

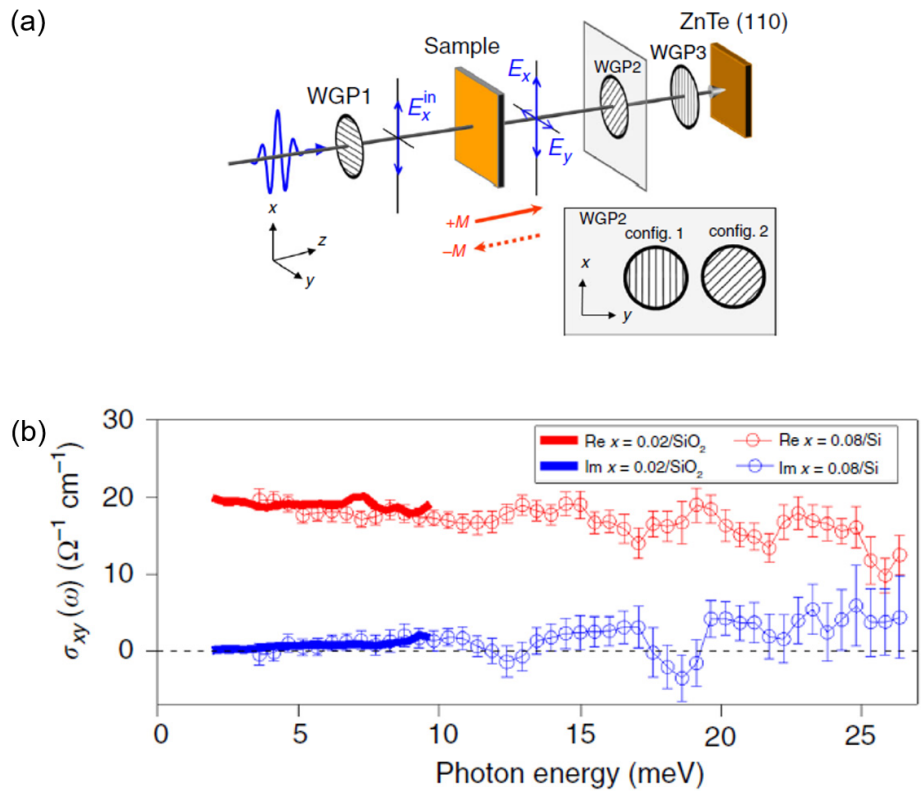

**Fig. 5** (a) The schematic of the polarization-resolved measurement setup. WGP denotes the wire-grid polarizer. (b) The real- and imaginary-part of the Hall conductivity ($\sigma_{xy}$) spectra for $Mn_{3+x}Sn_{1-x}$ films. The solid curves display the low-frequency THz-time-domain spectroscopy for $x = 0.02$ on a $SiO_2$ substrate while the open circles show the broadband spectrum for $x = 0.08$ on a Si substrate.[116] Copyright 2020, The Authors, published by Springer Nature.

In summary, the coexisting ferromagnetic and antiferromagnetic merits in noncollinear antiferromagnets have made them both intriguing in physics and promising for spintronic device applications. Consequently, noncollinear antiferromagnetic spintronics could soon become the center of focus for next-generation ultrafast and high-density information technologies. Moreover, for a more general perspective on noncollinear spintronics beyond antiferromagnets, a comprehensive review is available.[118]

*Note added.* After finishing this work, we realize that the perpendicular full electrical switching of the chiral antiferromagnetic order of $Mn_3Sn$ has been achieved.[119] In addition, strain-tunable spin wave[120] and oscillating-magnetic-field-tunable AHE[121] that is important for ultrafast optical pulse modulation have been theoretically revealed for noncollinear antiferromagnets $Mn_3X$.

**Acknowledgments:**

Z.L. acknowledges financial support from the National Natural Science Foundation of China (No. 52121001).